\documentclass{aastex701}

\usepackage{amsmath}
\usepackage{graphicx}

\begin{document}

\title{Identifying Casual SETI Transmissions and Ultra-Faint RFI with Cumulant Imaging}

\author[orcid=0000-0001-7694-4030,sname='Morales']{Miguel F. Morales}
\affiliation{University of Washington}
\affiliation{Dark Universe Science Center}
\email[show]{miguelfm@uw.edu}

\begin{abstract}
Techniques for identifying radio beacons transmitted by extra-terrestrial civilizations are well advanced. However, the casual broadband emission from standard radio communication is much harder to detect, and counter-intuitively casual Earth radio transmissions are becoming more difficult to identify as we more efficiently utilize the radio spectrum. In this paper we examine the electric-field statistics behind radio communication and re-conceptualize standard radio imaging as the first non-zero cumulant. This leads us to images of the higher cumulants where the natural stochastic emission from astrophysical sources disappears, leaving images of only modulated sources of radio communication. Faint SETI and RFI sources stand out in these cumulant images, enabling identification of faint communication that is  undetectable with current techniques. 
\end{abstract}


\section{Intro} 
\label{sec:intro}

Identifying faint artificial radio transmission is of interest both for Search for Extraterrestrial Intelligence (SETI) and for identifying and excising faint Radio Frequency Interference (RFI). A fundamental challenge for both radio SETI and ultra-faint RFI is distinguishing the artificial transmission from the ubiquitous natural radio emission. One approach is to look for spectral lines, particularly lines with no known astrophysical source. Another approach is to include the time domain to look for broadband flashes, chirps, or other  unnatural features in the time-frequency domain (see \citealt{NASAtechnosignature2019} for review). 

However, to make efficient use of the radio spectrum modern digital radio communications are spectrally smooth and steady in time. As our appetite for radio bandwidth has grown, the casual radio communication emission from the Earth has become harder to identify by a distant civilization as we more efficiently use the radio spectrum. The anthropogenic emission from the Earth is becoming spectrally smooth and steady. Identification of ultra-faint RFI faces the same challenge, where old analog broadcasts with strong pilot tones are much easier to identify than modern digital TV and cellphone transmissions with steady, broad, featureless spectra. 

However, the electric-field statistics of radio communications do not look like the stochastic emission of natural sources. When the electric field is viewed in a constellation diagram (a histogram of the complex electric field in a spectral channel) the distribution of various encoding schemes (e.g.\ phase shift keying, quadrature amplitude modulation, etc.) is statistically distinct from the Gaussian distribution of natural stochastic radio emission (e.g.\ thermal, synchrotron, free-free, natural emission lines, etc.). This separation is not accidental---it is used to distinguish the  communication signal from the background noise during demodulation. However, once the signal-to-noise ratio drops too low it becomes fundamentally impossible to demodulate the signal and demodulation searches for SETI and RFI fail. 

In this paper we ask a slightly different statistical question:  can we use the distinct electric-field statistics of radio communication to image artificial transmissions even when the signal-to-noise is too low to demodulate the signal? Much like angular Power Spectra allow cosmologists to identify patterns in the Cosmic Microwave Background (CMB) even when there is insufficient signal-to-noise to make an image; here we use statistics to identify radio communication even when the signal-to-noise is too low to demodulate the transmission. This statistical approach allows us to identify a class of signals that are fundamentally undetectable with current SETI and RFI identification techniques.

\begin{figure}[t]
    \centering
    \includegraphics[width = \linewidth, alt={Two images, the first contains a standard simulated starfield, where one of the objects also contains some modulated emission but it cannot be identified. In the other image based on the fourth cumulant all the natural stars disappear and only the modulated emission from one object remains.}]{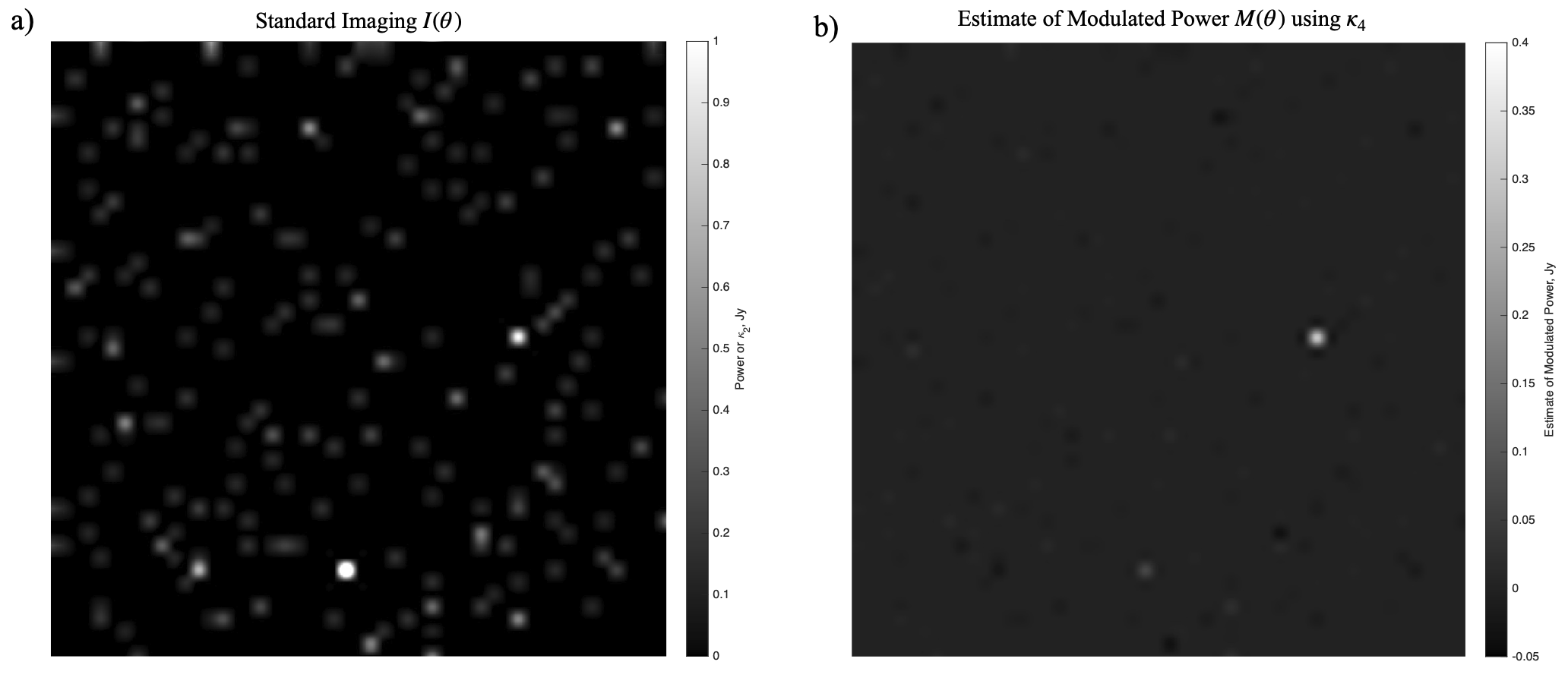}
    \caption{Simulated images from a star field where one of the stellar systems also includes a broadband radio broadcast. We simulate the full electric field in each direction for 8~seconds with 100~kHz bandwidth, with 200 natural sources and one stellar system with 0.7~Jy of natural emission and an additional 0.3~Jy from a phase modulated radio transmission. The lefthand image a) shows standard radio imaging, or the image of the second cumulant at each location $\kappa_2(\theta)$. As the transmission is broadband (no spectral features) and has too low an SNR to demodulate, this emission is \textbf{undetectable} using current SETI search techniques. Panel b) shows an image of the estimated modulated power $M(\theta)$ based on the negative signed square root of the fourth cumulant $\kappa_4(\theta)$. As the fourth cumulant is zero for natural stochastic emission the astrophysical sources all disappear in the righthand panel with small random fluctuations around zero. The source with modulated transmission stands out at greater than 
    $40\sigma$ and the 0.3~Jy of modulated power is correctly recovered (note change in color bar range). This can be repeated at each observed frequency channel to determine the full broadcast spectrum. Cumulant imaging allows a new approach for detecting the casual emissions from radio communication by other civilizations and for quickly identifying sources of radio interference. }
    \label{fig:images}
\end{figure}

The key result of the paper is shown in Figure~\ref{fig:images}. We simulated the electric field (8~sec, 100~kHz bandwidth) of a star field with 200 natural astrophysical sources and one source where 0.7~Jy of the emission is natural emission from the system (star, planet, and/or background synchrotron) and 0.3~Jy is from a broadband phase modulated radio broadcast. The lefthand panel shows the standard radio image, which is equivalent to an image of the second cumulant~$\kappa_2$. This casual emission cannot be demodulated (noise dominated) and has no spectral or temporal signature (steady broadband), so cannot be detected with current techniques. The righthand panel shows an image based on the the fourth cumulant~$\kappa_4$. In this image all natural stochastic emission vanishes, leaving only the 0.3~Jy of modulated communication.

This work expands on a long history of high-order electric-field statistics for RFI mitigation and SETI detection. For RFI identification in a receiver signal the spectral kurtosis is defined as the fourth cumulant \citep{Vrabie2003} and has been explored and used in a number of contexts \citep{Nita2007,TaylorCHIME2019,Smith2022,Tonder2023}. In this paper we project the electric field from the radio telescope to the sky to create an electric field image, then re-conceptualize imaging as a cumulant expansion of the electric field image. In this re-framing traditional radio imaging corresponds to the first non-zero cumulant of the electric field image. We then explore how anthropological and SETI radio communication would appear in the higher cumulant images. In this paper I will primarily concentrate on the SETI applications, but the general approach applies equally well to widefield imaging and detection of faint RFI sources.


The bulk of this paper develops the statistics that lead to the images in Figure~\ref{fig:images}. In \S\ref{sec:cumulants} we review the elegant mathematics of cumulants and why they are so powerful for this application. We follow this by analyzing the symmetries that emerge in the low SNR regime (\S\ref{sec:symmetries}) and how this greatly simplifies the statistics, and use this to calculate the cumulant signatures of common modulation schemes (Table~\ref{table}). We then develop cumulant imaging in \S\ref{sec:cumImaging}. Sections \S\ref{sec:sensitivity}~\&~\S\ref{sec:strength} then explore the sensitivity and scaling relations for cumulant imaging. We discuss coherent astrophysical sources such as pulsars and masers in \S\ref{sec:coherent}. We then conclude in \S\ref{sec:discussion} by discussing the steps needed to realize cumulant imaging and how cumulant images would complement current RFI and SETI searches.

\section{Cumulants}
\label{sec:cumulants}

When studying statistical distributions it is often useful to expand the probability density function $\rm{pdf}()$ in a series such as the familiar moments of a distribution: mean, variance, skew, kurtosis, etc. The cumulants $\kappa$ are closely related to the moments $\mu$ as an alternate expansion for statistical pdf, and cumulants have a couple of key features for our application:
\begin{itemize}
    \item \textbf{For zero-mean Gaussian probability distributions, only the second cumulant $\kappa_2$ is non-zero.} The electric-field for essentially all steady natural emission---whether it be thermal, synchrotron, free-free, or spectral line emission---is statistically drawn from a stochastic random Gaussian distribution. Physically the emission is the admixture of light emitted by an enormous number of independent micro-emitters such as many individual electrons shifting molecular levels (astrophysical line emission) or spiraling in magnetic fields (synchrotron). For stochastic sources the central limit theorem guarantees the pdf of the resulting electric-field will have a Gaussian (normal) distribution. The consequence is that all cumulants for a natural stochastic source are zero except for the variance, $\kappa_2$.
    \item \textbf{When independent random variables $B$ and $C$ add ($B + C = A$), the cumulants of the resulting random variable $A$ are the sum of the cumulants for $B$ and $C$.} This is the defining feature of cumulants (thus the name), and is why they are sometimes more useful than the moments. For our application it makes calculating the cumulants of composite sources straightforward \S\ref{sec:strength}.
    \item \textbf{The Gaussian distribution is the only distribution for which all higher cumulants are zero.} This implies that we should expect the pdfs of communication signals to have structure in their high order cumulants that is distinct from the zeros of stochastic astrophysical signals.
\end{itemize}

The pairing of cumulants adding and the high cumulants of stochastic astrophysical sources being uniquely zero make cumulant expansions of communication signals interesting to pursue.

\section{Low SNR symmetries}
\label{sec:symmetries}

The electric-field over a narrow bandwidth from a radio receiver is naturally expressed as a complex number $E$. The two-dimensional structure of complex plane has important consequences for the moments and cumulants, and makes them more complicated than one dimensional moments and cumulants. In particular there are many moments and cumulants at each order, and even for a single cumulant the number of terms can become enormous---for complex random variables the symmetric eighth cumulant has 109 terms. However, we can simplify our problem enormously by considering the symmetries of low SNR measurements.

These symmetries can be visually understood by considering the constellation diagram of the electric-field as shown in the first row of Table~\ref{table}. A constellation diagram is the pdf of $E$ in the complex plane (Electrical Engineers usually label the axes $I$ and $Q$ for in-phase and quadrature, while we'll use the physicist labeling of $r$ and $i$ for real and imaginary). In the first column the circularly symmetric two-dimensional constellation diagram of $E$ for a natural incoherent source is shown. 

\begin{table}[t]
    \centering
    \vspace{1 cm}
    \includegraphics[width=1\linewidth, alt={A table with a column for incoherent natural emission, phase modulated transmission, and 16 QAM modulated emission. The rows show a constellation diagram, parameters of the emission, and analytic values for the second, fourth, sixth, and eighth cumulants.}]{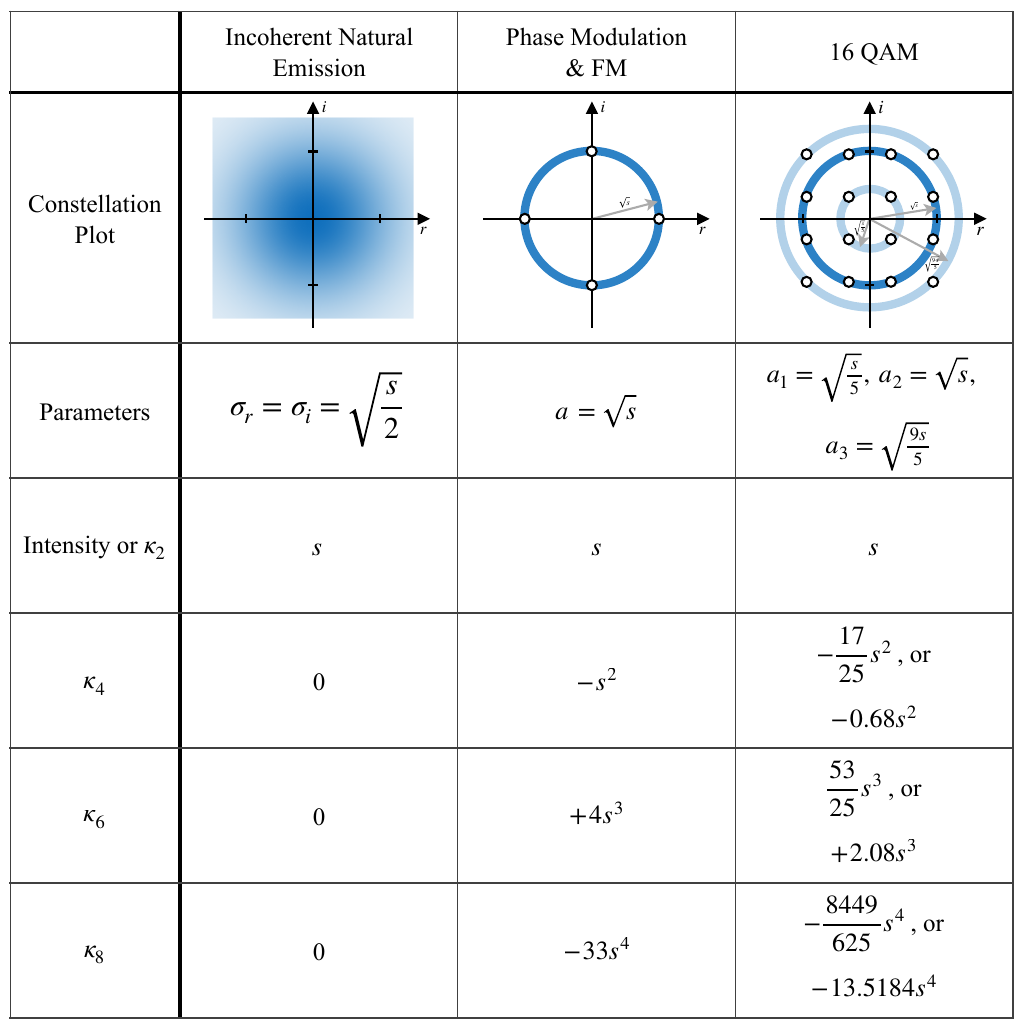}
    \caption{This table shows the statistical properties (left-to-right) of incoherent natural emission, phase modulated and FM signals, and 16 QAM modulated signals all normalized to the same intensity $s$. The first row shows a constellation plot of the E-field distribution, with the blue rings in phase and QAM indicating the low SNR distribution when there is no phase or clock lock. The second row indicates the associated parameters for the distributions, followed by cumulants $\kappa_2$ through $\kappa_8$. For incoherent natural emission only the second cumulant $\kappa_2$ exists and is equivalent to the brightness $I$ and variance of the electric field $\langle EE^*\rangle$. A Gaussian random process is fully described by the variance. All other statistical distributions have non-zero higher cumulants, with the analytical values for phase and 16 QAM modulated digital signals shown. The fourth cumulant $\kappa_4$ is particularly useful for identifying modulated transmissions, with the ratios of the higher cumulants being useful for identifying the modulation family. }
    \label{table}
\end{table}

The white circles in the second constellation diagram ($1,i,-1,-i$) indicate the encoding for quadrature phase shift keying (PSK). For quadrature PSK the transmitted electric field will move from one of the four locations to another, with each symbol encoding two bits. Over a long message the constellation diagram is the pdf of the transmitted electric-field consisting of four $\delta$-functions equally spaced around the origin. 

One of the challenges in demodulating a PSK signal is determining the initial phase and tracking the apparent frequency. The initial phase depends on the time delay of the signal, so the raw received electric field has an unknown phase rotation. This is often resolved by sending a known key sequence at the beginning of a packet to remove the phase ambiguity. In addition, the exact frequency of the transmitter and the receiver will differ due to Doppler shifts and clock errors. This leads to a slow rotation of the received electric-field constellation diagram, and this frequency offset must be corrected for to properly demodulate the signal.

But we are interested in the very low SNR regime. When the noise becomes high, we can no longer demodulate the signal (the symbols overlap due to convolving with Gaussian noise). This means that we cannot solve for the initial phase nor stop the phase rotation due to Doppler shifts and clock offsets. Due to these unknown rotations in the complex plane, the pdf of the transmitted electric-field becomes \textbf{circularly symmetric} at the observer as indicated by the blue ring. While we have used the example of a quadrature phase modulated signal, any phase or frequency modulation scheme with constant amplitude will have the same circular pdf shown by the blue ring in the low SNR regime.

The righthand column shows the constellation diagram for a 16 symbol Quadrature Amplitude Modulation (16 QAM) encoding scheme. Again, if the signal cannot be demodulated there are phase ambiguities that will rotate the received pdf. If the symbols are uniformly populated, the low SNR pdf consists of three rings with the central ring being twice as probable (8 symbols) as the inner and outer rings (4 symbols each).  

While we have used the PSK and QAM modulated signal as common examples, the key takeaway is that in the low SNR regime the pdf of the transmitted signal is circularly symmetric in the complex plane. The statistics of complex electric-field signals is deep and subtle, and exhaustively covered in the book by \cite{Schreier_Scharf_2010}. However, these expressions simplify enormously when, as in our case, the electric-field is a zero-mean circularly symmetric distribution (see \citealt{Maiello2023} for a nice discussion).

Crucially:
\begin{itemize}
    \item The moments can be expressed as $\mathbb{E}\left[Z^n Z^{*m}\right]$, where $\mathbb{E}\left[\ \right]$ is the expectation.
    \item All unbalanced ($n\neq m$) moments $\mu_{n,m}$ and  cumulants $\kappa_{n,m}$  are \textbf{zero}.
\end{itemize}
Conceptually we can calculate products of complex random numbers $Z$ and their conjugates $Z^*$, but if $Z$ is circularly symmetric there must be an equal number of normal and conjugated copies for the resulting product not to be circularly symmetric. Combined with the expectation of a circular distribution being zero by symmetry, this eliminates the vast majority of moments and cumulants. Below the first four non-zero cumulants are listed for the circularly symmetric zero-mean complex electric-field $E$ observed by a radio telescope: 

\begin{eqnarray} 
\kappa_2 &= \kappa_{1,1} = & \langle EE^*\rangle, \label{eq:k2}\\
\kappa_4 &= \kappa_{2,2} =& \left< (EE^*)^2 \right> - 2\langle EE^*\rangle^2, \label{eq:k4}\\
\kappa_6 &= \kappa_{3,3} =& \left< (EE^*)^3 \right> -9\left< (EE^*)^2 \right>\langle EE^*\rangle +12\langle EE^*\rangle^3, \label{eq:k6}\\
\kappa_8 &= \kappa_{4,4} =& \left< (EE^*)^4 \right> 
        -16\left< (EE^*)^3 \right>\langle EE^*\rangle 
        +144\langle EE^*\rangle^2\left< (EE^*)^2 \right>
        -18\left< (EE^*)^2 \right>^2
        -144\langle EE^*\rangle^4. \label{eq:k8}
\end{eqnarray}

The notation for complex cumulants is somewhat variable, with the $\kappa_{n,m}$ form being common and aligning with the Mathematica functions used to enumerate high cumulants. However, since only the balanced cumulants exist ($n=m$) and the resulting cumulants are real I've used the non-standard notation $\kappa_l$ where $l = n+m = 2n$ (listed first above). In this notation $l$ nicely counts the number of factors of $E$ in the estimator, with half of them conjugated, and lines up better with the language description associated with the more familiar one-dimensional moments and cumulants. 

Reading the equations above, the first non-zero cumulant $\kappa_2$ is the mean of $EE^*$. This is very familiar as the measured brightness of a source and equivalently the variance or second moment of the electric-field. All standard radio imaging consists of calculating $\langle EE^*\rangle$ as a function of direction on the sky.

The fourth cumulant $\kappa_4$ is the fourth moment of the electric field $\left< EEE^*E^* \right>$, or equivalently $\left< (EE^*)^2 \right>$, minus 2 times the square of the second moment $\langle EE^*\rangle$. The sixth and eighth cumulants $\kappa_6$ and $\kappa_8$ contain combinations of six and eight products of $E$ respectively. In reading these relations the order of taking a mean and a power is crucially important, e.g.\ the square of the average is not equal to the average of the squares and the moments need to calculated at the sample rate. 

We can use these relationships to calculate the cumulants for each signal in Table~\ref{table}, where we have normalized all of the signals to have the same brightness $s$. The first row below the constellation diagram lists the parameters of the normalized signal, with the following rows listing the cumulants $\kappa_2$--$\kappa_8$. Each column then shows the cumulants for different types of signal. As expected for a stochastic astrophysical source $\kappa_2$ is the brightness $s$ and all higher cumulants are zero. 
For the PSK signal (or any signal in the phase-modulation or narrow frequency modulation family) the fourth cumulant is minus the brightness squared ($-s^2$). Conceptually the minus sign on the fourth PSK cumulant is because the amplitude of a PSK signal is steady and does not have the large outliers expected of a Gaussian distribution. The sixth cumulant is positive and proportional to the brightness cubed, while the eighth cumulant is again negative and proportional to the brightness to the fourth power. 
The 16 QAM signal follows the same positive-negative and powers of brightness sequence as the PSK signal, but with different pre-factors and ratios between the cumulants. The cumulants of other modulation schemes can be calculated using the simple ring of PSK as a template and averaging by the fractional occupations. 

More generally all existing modulation schemes have non-zero higher order cumulants. This is not an accident. To send radio communication it is important to be distinct from the ubiquitous thermal and astrophysical noise. In a companion paper Morales \& Smith (\textit{in review}), instead of trying to detect radio communication we ask if it is possible to hide radio communication. The conclusion of that paper is that it is possible to perfectly hide radio communication, but it places very strict constraints on the transmission scheme---constraints no current anthropogenic radio communication meets and are impractical for common radio communication needs. All practical radio communications have non-zero higher cumulants, while these cumulants are zero for all stochastic astrophysical emission.

\section{Cumulant Imaging}
\label{sec:cumImaging}

We can re-frame radio imaging as a cumulant expansion in each direction, with standard radio imaging being just the second cumulant $\kappa_2(\theta)$ measured at each location on the sky. While this re-framing may feel like mathematical overkill, it is interesting when searching for communication signals with non-zero higher cumulants. 

There is a class of new and near-future radio instruments that internally form the electric-field image of the sky $E(\theta)$. Inspired by the scientific promise of high time resolution science (e.g.\ Fast Radio Bursts) and the computational challenge of cosmology radio telescopes with many thousands of antennas there has been the development of radio correlators and beamformers that internally create widefield electric field images such as the CHIME FRB backend \citep{CHIMEFRB2018} and the MOFF/EPIC correlator on the LWA \citep{Morales2011,LWAEPIC2023}. Internally these `imaging correlators' form an electric-field image of the sky $E(\theta)$, then calculate the brightness $\langle EE^*\rangle(\theta)$ over short periods of time for each location on the sky. Mathematically these imaging correlators are calculating the second cumulant of the electric-field image to characterize the observed sky. Similarly the imaging focal plane receivers on the Five-hundred-meter Aperture Spherical radio Telescope (FAST; \citealt{FAST2020}), the Green Bank Telescope (GBT, \citealt{ARGUS2014}), the Parkes and Effelsberg telescopes \citep{CSIRO_PAF2016}, and the Australian Square Kilometre Array Pathfinder (ASKAP; \citealt{ASKAP2009}) use analog radio optics to form electric-field images. It would be straightforward to update instruments that form electric-field images to calculate higher moments such as $\langle (EE^*)^2\rangle(\theta)$ in addition to the standard $\langle EE^*\rangle(\theta)$ to enable imaging of higher cumulants. 

The reason this has never been pursued is that the higher order cumulants are zero for nearly all astronomical sources. The electric-field from astronomical sources is assumed to be `noise-like' (Gaussian distributed), so forming the higher moments and cumulants is pointless---for astronomical sources all of the information is contained in $\langle EE^*\rangle(\theta)$. 

However, these higher order cumulants are not zero for RFI and any SETI source that uses Earth-like radio communication. These sources will appear strongly in $\kappa_4(\theta)$, $\kappa_6(\theta)$, $\kappa_8(\theta)$ maps. We simulated this approach in Figure~\ref{fig:images} with 200 natural stochastic sources and one source with 0.7~Jy of natural emission and 0.3~Jy from a PSK modulated transmission. The lefthand image is a standard radio image $\kappa_2(\theta)$. We could have made an image of $\kappa_4(\theta)$ where the PSK source would have shown up as strong negative with units of Jy$^2$. For illustration clarity we instead formed a simple estimator of the modulated power $M(\theta)$ by taking the negative~signed~square root of $\kappa_4(\theta)$. This image has awkward statistics (see \S\ref{sec:sensitivity}), but is in positive Jy units that are easy to visualize. In this image all of the astrophysical emission disappears (mean zero with fluctuations), including the 0.7~Jy of natural emission at the location of transmitter. What is left is just the location and brightness of the radio communication. This makes identifying casual SETI or anthropogenic radio transmissions straightforward.  

This $M(\theta)$ modulated power estimator presumes that the signal is using phase or FM modulation, though the power from a 16~QAM source would have only been underestimated by $\sim$18\%  ($\sqrt{0.68}$ from Table~\ref{table}, third column). This modulated power estimator is in no way optimal, and in future work we intend to explore improved estimators of the modulated power using the cumulants.

Our analysis so far has focused on detecting radio communications in a single frequency channel. The analysis can be performed at all measured frequency channels to form a spectrum of the radio communication $M(f)$. In any SETI or RFI search identifying and measuring the spectrum of the artificial transmission can provide key information about the source. Lastly, the ratios of higher cumulants can provide information about the family of encoding being used at each frequency. Even when we don't have the SNR to demodulate the signal, the cumulants can isolate the modulated signal, measure its spectrum, and provide information about the encoding. 








\section{Sensitivity}
\label{sec:sensitivity}

In this section we start to explore the sensitivity of cumulant searches, and the characteristics of cumulant sensitivity are different from the power sensitivity (e.g. radiometer equation) that we as radio astronomers are familiar with. The conclusion is that whether traditional power measurements or cumulant searches are more sensitive for identifying a particular SETI or RFI signal depends intimately on the scenario considered. For a narrow line beacon---a common target for SETI searches---current search techniques are more sensitive than a cumulant search. However, broadband casual emission such as digital TV, cell phone, or communicating with interplanetary probes range from very hard to formally impossible to separate from natural astrophysical emission using standard search techniques and work very well using cumulants. 

We start by studying the statistics of cumulant estimators. For a null hypothesis---a source is natural and not an RFI or SETI source---the key questions are associated with the fluctuations of the cumulant estimate around zero when the source is a natural stochastic Gaussian random variable. In any location on the sky the noise can be characterized by the pixel noise power $s_N$ (Jy/beam or similar units), which is a combination of the instrument noise projected to the sky, diffuse sky emission (e.g.\ galactic synchrotron, CMB), and compact astrophysical emission such as the natural emission from the stellar system being observed. 

\begin{figure}[t]
    \centering
    \includegraphics[width=1\linewidth, alt={Two panels, one showing the nearly perfect Gaussian distribtuon of the fourth cumulant, and the other showing a bi-modal distribution of the modulated power estimate.}]{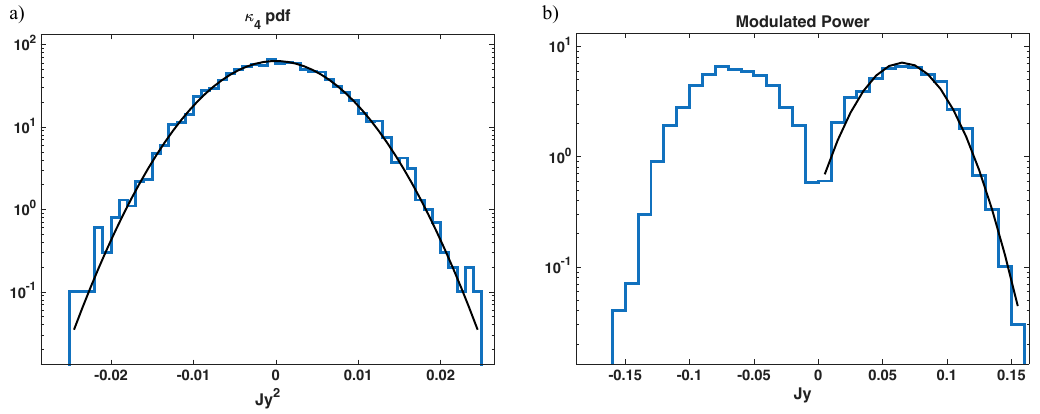}
    \caption{Simulated distributions from 10 thousand stochastic Gaussian sources with a brightness of 1~Jy and bandwidth time product $Bt = 1 \times 10^5$. The distribution of the fourth cumulant shown in panel a) is approximately normal (Equation~\ref{eq:ksig}, black line). The distribution of the Modulated Power estimate in panel b) is more complicated, both being bi-modal and not well approximated by a normal distribution (black line fit for positive values). See text for detailed discussion.}
    \label{fig:histograms}
\end{figure}

From detailed simulations, the fluctuations in the fourth cumulant estimate (Figure~\ref{fig:histograms}a) follow an approximately Gaussian distribution with
\begin{equation}
    \sigma_{\kappa4} = \frac{2 s_N^2}{\sqrt{Bt}}, 
     \label{eq:ksig}
\end{equation}
where $Bt$ is the bandwidth time product or the number of complex samples.

I do not have a full analytic description of the cumulant pdf, and since the simulations must start with the electric-field it is difficult to create enough realizations to study the tails of the pdf (they might be slightly heavy?).  Looking at the components of $\kappa_4$ (Equation~\ref{eq:k4}), the distribution of the first term $\left< (EE^*)^2 \right>$ is a Weibull distribution with shape parameter $k=1/2$ and scale parameter $\lambda = s_N^2$, leading to a standard deviation of $\sigma = \sqrt{20}\, s_N^2/\sqrt{Bt}$. The second term $2 \left<EE^* \right>^2$ has a non-central chi-squared distribution with $\sigma = 4 s_N^2/\sqrt{Bt}$. However, the fluctuations of these terms are strongly correlated and the variance of the cumulant $\kappa_4$ is significantly narrower than the two terms. (Equation~\ref{eq:ksig} can be analytically confirmed using Wick's theorem, Matthew McQuinn \textit{personal communication}.) Understanding the full non-Gaussian statistical significance (particularly when accounting for large trials factors) will necessitate an analytic understanding of the cumulant pdf and is left for future work.


It is tempting to use the modulated power estimate $M(\theta)$ (signed square root of $\kappa_4$) for data analysis. However, as shown in Figure~\ref{fig:histograms}b) the pdf of the modulated power estimate is bimodal. Even if we limit ourselves to one sign of fluctuations, the distribution is not well approximated by a normal distribution (overlay). After some study, I think the correct approach is to perform statistical tests in the native cumulant images and reserve this modulated power estimate for visualization and interpretation. 

                                            Returning to our estimate of the fourth cumulant pdf in Equation~\ref{eq:ksig} we can study the key sensitivity scaling relations of cumulant searches. While it is tempting to see the $\sqrt{Bt}$ in the denominator and expect it to scale like a familiar power search (e.g. the radiometer equation), the amplitude of the $\kappa_4$ cumulant signal (Table~\ref{table}) and the amplitude of the fluctuations scale as $s^2$ and $s_N^2$ respectively. The units of $\kappa_4$ are brightness squared not the brightness, and halving the modulated power reduces the fourth cumulant amplitude by a factor of 4. Consequently while for traditional power detection improving the threshold power sensitivity by a factor of 2 requires increasing the bandwidth time product by a factor of 4, for a cumulant search using $\kappa_4$ improving the sensitivity to the modulated power $s$ by a factor of 2 requires increasing the bandwidth time product by a factor of 16. It is the ratio between the power of the numerator and the power of the denominator that sets the scaling. The scalings for $\kappa_6$ and $\kappa_8$ are even higher.

But this scaling should not be over-interpreted either. In our simulation (Figure~\ref{fig:images}) 0.3~Jy modulated power was detected at more than $40\sigma$ in 8~seconds of observing at 100~kHz despite a 0.7~Jy of natural background emission from the stellar system. Part of the reason for this sensitivity is that the astrophysical background has already been subtracted by the cumulant estimator. Much like hunting for transients in circular polarization is often much more sensitive than searches for unpolarized emission due to the sensitivity penalty of foreground removal, the sky is dark for cumulant images. Accurately disentangling transmission from the background is part of what has driven SETI searches to look for narrow emission technosignatures. As background removal is already done in a cumulant search, much broader bandwidths can be used. For RFI searches the full bandwidth of an allocation (often several MHz) can be used, and if distant civilizations are efficiently using the radio spectrum the bandwidth of the entire water window could be used for a detection with cumulants.

If a distant civilization is trying to communicate with us by transmitting a distinctive power signal---such as a spectral line or pulsed transient---then the most sensitive way to detect this signal is to look in power vs.\ frequency and time and current SETI searches are optimally tuned. However, if the transmitted signal does not have a distinctive spectral or temporal fingerprint the statistical question changes. In a scenario with steady broadband transmission we can no longer use power in frequency and/or time to isolate the artificial transmission from natural emission. On Earth as we have more efficiently used the radio spectrum terrestrial radio communication has become less distinctive in the spectrum---it is approaching a featureless broadband signal. However, the natural radio background is a challenge for all forms of radio communication and the electric field of modulated radio transmissions are purposefully non-Gaussian to facilitate communication. Cumulants are tuned to look for this statistical feature of radio communication. Cumulants do not compete with standard SETI searches, they instead allow us to explore a different scenario where the transmitter is efficiently using the radio spectrum for communication.




\section{Predicting Cumulant Strength}
\label{sec:strength}

The mathematical features of cumulants make it surprisingly easy to predict the expected strength of a signal in various scenarios.
Cumulants are defined so that the when statistically independent signals are added the cumulants also add. This makes predicting the cumulants of a composite signal very straightforward. Two illustrative examples are how the cumulant signal changes when additional transmitters are added compared to when a total transmission power is sub-divided between independent transmitters. 

The first scenario starts with our simulation (Figure~\ref{fig:images}) where we have a natural source and a single transmitter in the channel of interest. As the second cumulant $\kappa_2$ is equivalent to the power, the power of the natural source and the transmitter simply add as we expect. Looking at higher cumulants $\geq\kappa_4$ the cumulants of the natural emission are zero, so the cumulants of the composite signal are equal to the cumulants of the transmitter (Table~\ref{table}). If we then add a second transmitter all of the cumulants just add again: the total power increases by the added transmission power and the amplitude of each cumulant is simply the sum of the cumulants of the two transmitters. Two transmitters in the same channel does cause interference and is actively regulated on Earth. But from outside the solar system  often a large number of transmitters will be simultaneously visible and the strength of the cumulant signal will be the sum of the cumulants of the individual transmitters. 

This scaling is nice:  the more radio communication the stronger the cumulant signal. However the $s^2$ scaling of $\kappa_4$ preferences strong transmitters. If two transmitters of equal power are added, the cumulants double. But if a single transmitter doubled its power the fourth cumulant would quadruple. So while the cumulants add, if we understand transmitters by their power $s_t$, when we are observing many overlapping transmissions the strength of the cumulant is primarily associated with power of the strongest typical transmitters, with weaker transmitters contributing relatively little.

While for RFI detection the number of overlapping transmitters is small, for SETI it is sometimes convenient to consider the aggregate transmission power from a civilization ($s_{\rm civ}$). If this aggregate power is due to $N$ independent transmitters the cumulant amplitudes \textit{decrease} as the total power is distributed over more transmitters $N$, with $\kappa_4 \propto s_{\rm civ}^2/N$. Conceptually this is expected as when we average many independent random variables the statistical distribution becomes increasingly Gaussian, and Gaussian distributions are uniquely defined by all higher cumulants being zero. This observation that all cumulants greater than $\kappa_2$ decrease when one averages larger numbers of random variables is at the center of some proofs of the central limit theorem. 

Looking at the Earth from afar, the strength of the cumulant signal is associated with the typical power of strong transmitters. Even if the aggregate transmitted signal in digital TV and cell bands was the same, since the fourth cumulant is proportional to $Ns_t^2$ the digital TV signal from a moderate number of strong broadcasters would be stronger than the signal from a large number of weaker cell phone towers. For predicting the cumulant signal in a SETI context it is the number and power of the brightest transmitters that matters.

\section{Coherent Astrophysical Sources}
\label{sec:coherent}

Most astrophysical emission is incoherent:  the light is generated by the stochastic emission from a very large number of independent electrons, and the central limit theorem guarantees this emission will be highly Gaussian. The classical and quantum theory of this emission and the two and four point statistics is discussed in \cite{Hirata2014}. However, there are coherent astrophysical sources such as pulsars and masers, and strong modulation by scintillation can also potentially produce non-zero higher cumulants. The observational measurements of high order statistics are unfortunately rather thin, but the observations we do have imply that the negative  fourth cumulant ($\kappa_4$) produced by radio communications (Table~\ref{table}) is likely to be unique. 

\cite{Tonder2023} used tied array MeerKAT observations of the Vela pulsar to measure the spectral kurtosis for RFI excision, and finds that that for long estimation windows that the pulsar has excess positive kurtosis while the RFI has negative kurtosis as expected from Table~\ref{table}. Conceptually sources with bright coherent flashes are expected to have larger outliers than one would expect given their average power, leading to a positive fourth cumulant. While not zero as for incoherent astrophysical signals, this positive pulsar $\kappa_4$ is distinct from the negative fourth cumulant from radio communications. The statistical signature of RFI and SETI signals is a deficit of large outliers. Similarly I would expect strong scintillation of an incoherent background source to produce a positive $\kappa_4$.

Theoretically a maser can have a negative $\kappa_4$ if the amplification saturates, clipping the amplitude produced from incoherent seed photons. \cite{Evans1972} measured the four point statistics of a number of OH masers with the National Radio Astronomy Observatory 140~ft telescope and found no deviations from Gaussianity greater than 1\%. As maser emission occurs at only a few well known frequencies, even if some sources do produce a negative $\kappa_4$ it would not be a significant foreground contaminant for a SETI search. 

This is not a comprehensive catalog of astrophysical foregrounds and natural sources with negative $\kappa_4$ are still a significant concern, however it must be produced by a coherent process with a deficit of large outliers in the electric-field magnitude.

\section{Discussion}
\label{sec:discussion}

In this paper we have re-conceptualized radio observations as a series of cumulants of the electric field image, where traditional radio imaging uses the first non-zero cumulant $\kappa_2$. As it is only this second cumulant that exists for stochastic astrophysical signals, it makes sense to simply measure the power and conceptualizing imaging as a cumulant expansion is mathematical overkill. However, higher order cumulants are useful for isolating the modulated emission from artificial sources and re-framing as a series puts traditional imaging and cumulant imaging on the same mathematical footing, allowing us to more easily compare the pros and cons of various analysis approaches. 

My current view is that traditional ($\kappa_2$) and higher order cumulant imaging ($\kappa_4$ and above) are complimentary for detecting radio RFI and SETI. If an alien civilization wants to be seen, the most power efficient approach is to broadcast a distinctive power beacon in frequency and/or time. This will give the greatest range for a given transmit power, and standard SETI searches are well optimized to detect these beacons. 

In the twentieth century casual radio communications from Earth looked a lot like beacons---we concentrated a lot of power in distinct non-physical lines with narrow modulation. The distinction between a beacon and casual emission was mostly a difference in transmitted power not the spectral features. However, our appetite for communicating ever more data over limited radio spectrum has made modern radio communications much harder to detect by a distant civilization. Modern digital transmissions are spectrally white over broad bands, with much of the narrow emission features being due to legacy protocols. It is not unreasonable to think that in another century the casual radio emission from the Earth will be nearly uniform across the water window (the frequencies that escape our atmosphere), making differentiating our technological radio emission from natural emission much more difficult. An advanced civilization efficiently using the radio spectrum might be invisible to us unless they were actively broadcasting a beacon. 

However, the statistics of radio communications are purposefully non-Gaussian, and cumulant imaging offers a mathematical framework for searching for casual broadband radio communications. An illustrative example is when the transmitted signal is steady, broadband, and cannot be demodulated. While a receiving telescope with larger collecting area and higher angular resolution can increase the gain of a received signal, it can only be demodulated if there is only one transmitted signal (no interference that originates at the source) and the signal power is greater than the noise from the source. A planet with a handful of broadcasters in the same band (different parts of the planet and/or stellar system), or a system where the broadcast power is fainter than the natural emission from the star/planets/local environment is effectively impossible to demodulate (requires resolving individual transmitters with SNR $>$ 1). Broadband, overlapping transmissions of moderate power is a good model for much of the casual radio emission from Earth.

It is in this scenario that cumulant imaging could be very effective. All Earth-like communications have non-zero higher cumulants, and the cumulants of interfering transmitters add---increasing their detectability in cumulant images. Further, as stochastic astrophysical sources have zero higher cumulants, the high order cumulant images isolate the modulated transmission from the sea of natural radio emission (Figure~\ref{fig:images}). While high order cumulants are less sensitive than a power search when looking for a beacon, they are much more sensitive when looking for Earth-like casual radio communications. Cumulant imaging offers the possibility of identifying the casual radio communications from a distant civilization and measuring their spectrum allocation, even if it is impossible to demodulate the signal and there is no spectral or temporal signature in the power emitted.  

While I have concentrated on the example of SETI,  widefield cumulant imaging would be exquisitely sensitive for localizing faint RFI sources such as satellite transmissions, reflections of digital TV from satellites and aircraft, ionospheric ducting, etc. One of the key challenges for identifying ultra-faint RFI is separating the transmission from the astrophysical background, with much of the most sensitive work being done in the context of removing RFI from 21~cm cosmology measurements (\citealt{Wilensky2019,Wilensky2020,Wilensky2023b,Kunicki2024}, Ducharme et al. \textit{in review}, Lilleskov et al. \textit{in prep.}). These efforts all use some statistical handle such as source motion relative to the sky or temporal power variation to disentangle the RFI from the astrophysical signal, and this background subtraction significantly impacts the sensitivity of these approaches. Cumulant imaging offers a new way to identify these RFI sources where the Gaussian astrophysical signal is naturally zero, potentially leading to much more sensitive RFI searches. Further, if the modulation scheme is known the brightness of the higher order cumulants  provide an accurate estimate of the RFI power, potentially enabling ways of subtracting the RFI and not just flagging it. While SETI is more intellectually entertaining, the real scientific promise of cumulant imaging may be in imaging and removing anthropogenic sources of RFI in sensitive radio science observations.

\section{Conclusion}
\label{sec:conclusion}

In this paper we expanded on the work by \cite{Vrabie2003} and others to re-conceptualize images of the sky as one term in a cumulant expansion. In this expansion traditional imaging is the first non-zero cumulant, and is the only cumulant that exists for stochastic astrophysical sources. However, the electric-field statistics of radio communication is purposefully distinct from stochastic noise and shows up in the higher cumulants. In this paper we explored the properties and symmetries of cumulants for radio observations (\S\ref{sec:cumulants}--\S\ref{sec:symmetries}, Table~\ref{table}), the properties of cumulant images (\S\ref{sec:cumImaging}, Figure~\ref{fig:images}), and the resulting statistical properties (\S\ref{sec:sensitivity}) and signal strength (\S\ref{sec:strength}). These higher order cumulant images may be particularly useful for identifying faint RFI and the casual radio emission of distant civilizations. 

However, while this paper provides the mathematical foundation for cumulant imaging there are large number of questions that will need to be resolved if cumulant imaging is to be used in practice. Here I briefly discuss a few common concerns and areas of necessary future work.

\textbf{Is a spectral cumulant survey technically possible?}  Yes. With the advent of imaging correlators and beamformers with thousands of spectral channels over broad bands, it is technically feasible to make spectral cumulant surveys. These remarkable digital machines internally generate electric-field image cubes with the electric field in hundreds to millions of directions (pixels) and thousands of fine frequency channels. Currently they calculate the power ($\kappa_2$) in each direction and frequency, but it would be a straightforward enhancement to calculate the fourth moment as well and possibly the sixth and eighth. This is sufficient to then form higher order cumulant spectral images (Equations~\ref{eq:k4}--\ref{eq:k8}). Most astrophysical sources should disappear in these kigher order cumulant images, while casual radio emission will stand out with a spectrum of the broadcasts. A non-Earth like spectrum of radio communication would be a strong indicator of SETI. 


\textbf{Are there astrophysical sources in the cumulant images?} Maybe. A key assumption in this paper is that the majority of astrophysical emission is stochastic with zero higher cumulants. Due to the microphysics this is known to be true for incoherent sources such as thermal emission, atomic and molecular emission, synchrotron emission, and free-free emission. There are a few observations of the fourth cumulant for coherent emitters as discussed in \S\ref{sec:coherent}, but I do not know of a comprehensive survey of the electric-field statistics of coherent astrophysical sources. It could be that a cumulant survey for SETI would be overwhelmed by non-stochastic astrophysical sources I have not considered here. Conversely, if these sources are scientifically interesting this could be a new way of finding and studying them. 

\textbf{Are there instrumental contaminants?} Almost certainly. The community has spent decades optimizing the analog and digital performance of radio instruments for power measurements. From amplifier linearity to van~Vleck corrections, how instruments can affect the second cumulant have been extensively studied and perfected. In addition there is a deep tradition of analysis methods to mitigate instrumental imperfections, such as differential radio photometry to correct gain changes for transient searches. There are almost certainly instrumental effects for higher order cumulant measurements that will need to be identified and corrected. For example, the digital non-linearity identified in \cite{Byrne2026} creates small spectral distortions important for radio cosmology measurements, but creates large broadband distortions in $\kappa_4$. Performing a real world cumulant survey will require significant investment in both instrumentation and analysis methods to achieve science quality observations.

In this paper I concentrated on instruments that can form an electric field image via radio optics, an imaging correlator, or a massively parallel beamformer. It is also possible to form interferometric cumulant images using a variation of a correlator. In an upcoming paper I explore interferometric cumulant imaging (Morales \textit{in prep.}) and more deeply explore the point spread function (array beam) of cumulant images. Computing the cumulant equivalents of interferometric visibilities scales poorly, but it can enable  high resolution VLBI imaging of distant radio transmissions and the use of deconvolution and other common interferometric data analysis techniques. 


\begin{acknowledgments}

Nichole Barry and Matthew McQuinn offered deep and insightful comments that greatly impacted the final paper. Nichole Barry framed the use case for casual SETI transmissions, and both Matthew McQuinn and Nichole Barry led key discussions of coherent astrophysical emitters. I'd also like to thank Joshua Smith, Bryna Hazelton, Nithyanandan Thyagarajan, and Daniel Jacobs for their help as I developed this concept.

\end{acknowledgments}

\begin{contribution}
All work and errors performed by the author. No AI was used. Simulations and scripts available on reasonable request. 
\end{contribution}

\appendix

\section{Raw Cumulants}

As convenience for the referee the $n=m$ complex cumulants in terms of the raw moments, generated using Mathematica. Unless advised otherwise I'd remove from a published version of the paper. These relations simplify to Equations~\ref{eq:k4}--\ref{eq:k8} when either the pdf is circularly symmetric or if only the circularly symmetric terms are kept. 





\begin{align} \kappa_{2,2} = &-6 \mu _{1,0}{}^2 \mu _{0,1}{}^2+2 \mu _{2,0} \mu _{0,1}{}^2+8 \mu _{1,0} \mu _{1,1} \mu _{0,1}-2 \mu _{2,1} \mu _{0,1}+2 \mu _{0,2} \mu _{1,0}{}^2  
 -2 \mu _{1,1}{}^2-2 \mu _{1,0} \mu _{1,2}\nonumber\\
 &-\mu _{0,2} \mu _{2,0}+\mu _{2,2}
\end{align}

\begin{align}
    \kappa_{3,3} = & -120 \mu _{1,0}{}^3 \mu _{0,1}{}^3+72 \mu _{1,0} \mu _{2,0} \mu _{0,1}{}^3- 6 \mu _{3,0} \mu _{0,1}{}^3+216 \mu _{1,0}{}^2 \mu _{1,1} \mu _{0,1}{}^2-54 \mu _{1,1} \mu _{2,0} \mu _{0,1}{}^2 \nonumber\\
    & -54 \mu _{1,0} \mu _{2,1} \mu _{0,1}{}^2+6 \mu _{3,1} \mu _{0,1}{}^2+72 \mu _{0,2} \mu _{1,0}{}^3 \mu _{0,1}-108 \mu _{1,0} \mu _{1,1}{}^2 \mu _{0,1}-54 \mu _{1,0}{}^2 \mu _{1,2} \mu _{0,1}\nonumber\\
    &-54 \mu _{0,2} \mu _{1,0} \mu _{2,0} \mu _{0,1}+18 \mu _{1,2} \mu _{2,0} \mu _{0,1}+36 \mu _{1,1} \mu _{2,1} \mu _{0,1}+18 \mu _{1,0} \mu _{2,2} \mu _{0,1}+6 \mu _{0,2} \mu _{3,0} \mu _{0,1}\nonumber\\
    &-3 \mu _{3,2} \mu _{0,1}-6 \mu _{0,3} \mu _{1,0}{}^3+12 \mu _{1,1}{}^3-54 \mu _{0,2} \mu _{1,0}{}^2 \mu _{1,1}+36 \mu _{1,0} \mu _{1,1} \mu _{1,2}+6 \mu _{1,0}{}^2 \mu _{1,3}+6 \mu _{0,3} \mu _{1,0} \mu _{2,0}\nonumber\\
    &+18 \mu _{0,2} \mu _{1,1} \mu _{2,0}-3 \mu _{1,3} \mu _{2,0}+18 \mu _{0,2} \mu _{1,0} \mu _{2,1}-9 \mu _{1,2} \mu _{2,1}-9 \mu _{1,1} \mu _{2,2}-3 \mu _{1,0} \mu _{2,3}-\mu _{0,3} \mu _{3,0}\nonumber\\
    &-3 \mu _{0,2} \mu _{3,1}+\mu _{3,3}
\end{align}


\begin{align}
    \kappa_{4,4} = &-5040 \mu _{1,0}{}^4 \mu _{0,1}{}^4-360 \mu _{2,0}{}^2 \mu _{0,1}{}^4+4320 \mu _{1,0}{}^2 \mu _{2,0} \mu _{0,1}{}^4-480 \mu _{1,0} \mu _{3,0} \mu _{0,1}{}^4+24 \mu _{4,0} \mu _{0,1}{}^4\nonumber\\
    &+11520 \mu _{1,0}{}^3 \mu _{1,1} \mu _{0,1}{}^3-5760 \mu _{1,0} \mu _{1,1} \mu _{2,0} \mu _{0,1}{}^3-2880 \mu _{1,0}{}^2 \mu _{2,1} \mu _{0,1}{}^3+576 \mu _{2,0} \mu _{2,1} \mu _{0,1}{}^3\nonumber\\
    &+384 \mu _{1,1} \mu _{3,0} \mu _{0,1}{}^3+384 \mu _{1,0} \mu _{3,1} \mu _{0,1}{}^3-24 \mu _{4,1} \mu _{0,1}{}^3 +4320 \mu _{0,2} \mu _{1,0}{}^4 \mu _{0,1}{}^2-8640 \mu _{1,0}{}^2 \mu _{1,1}{}^2 \mu _{0,1}{}^2\nonumber\\
    &+432 \mu _{0,2} \mu _{2,0}{}^2 \mu _{0,1}{}^2-216 \mu _{2,1}{}^2 \mu _{0,1}{}^2-2880 \mu _{1,0}{}^3 \mu _{1,2} \mu _{0,1}{}^2-4320 \mu _{0,2} \mu _{1,0}{}^2 \mu _{2,0} \mu _{0,1}{}^2\nonumber\\
    &+1728 \mu _{1,1}{}^2 \mu _{2,0} \mu _{0,1}{}^2+1728 \mu _{1,0} \mu _{1,2} \mu _{2,0} \mu _{0,1}{}^2+3456 \mu _{1,0} \mu _{1,1} \mu _{2,1} \mu _{0,1}{}^2+864 \mu _{1,0}{}^2 \mu _{2,2} \mu _{0,1}{}^2\nonumber\\
    &-216 \mu _{2,0} \mu _{2,2} \mu _{0,1}{}^2+576 \mu _{0,2} \mu _{1,0} \mu _{3,0} \mu _{0,1}{}^2-144 \mu _{1,2} \mu _{3,0} \mu _{0,1}{}^2-288 \mu _{1,1} \mu _{3,1} \mu _{0,1}{}^2-144 \mu _{1,0} \mu _{3,2} \mu _{0,1}{}^2\nonumber\\
    &-36 \mu _{0,2} \mu _{4,0} \mu _{0,1}{}^2
    +12 \mu _{4,2} \mu _{0,1}{}^2-480 \mu _{0,3} \mu _{1,0}{}^4 \mu _{0,1}+2304 \mu _{1,0} \mu _{1,1}{}^3 \mu _{0,1}-72 \mu _{0,3} \mu _{2,0}{}^2 \mu _{0,1}\nonumber\\
    &-5760 \mu _{0,2} \mu _{1,0}{}^3 \mu _{1,1} \mu _{0,1}+3456 \mu _{1,0}{}^2 \mu _{1,1} \mu _{1,2} \mu _{0,1}
    +384 \mu _{1,0}{}^3 \mu _{1,3} \mu _{0,1}+576 \mu _{0,3} \mu _{1,0}{}^2 \mu _{2,0} \mu _{0,1}\nonumber\\
    &+3456 \mu _{0,2} \mu _{1,0} \mu _{1,1} \mu _{2,0} \mu _{0,1}-864 \mu _{1,1} \mu _{1,2} \mu _{2,0} \mu _{0,1}-288 \mu _{1,0} \mu _{1,3} \mu _{2,0} \mu _{0,1}
    +1728 \mu _{0,2} \mu _{1,0}{}^2 \mu _{2,1} \mu _{0,1}\nonumber\\
    &-864 \mu _{1,1}{}^2 \mu _{2,1} \mu _{0,1}-864 \mu _{1,0} \mu _{1,2} \mu _{2,1} \mu _{0,1}-432 \mu _{0,2} \mu _{2,0} \mu _{2,1} \mu _{0,1}-864 \mu _{1,0} \mu _{1,1} \mu _{2,2} \mu _{0,1}\nonumber\\
    &+144 \mu _{2,1} \mu _{2,2} \mu _{0,1}-144 \mu _{1,0}{}^2 \mu _{2,3} \mu _{0,1}+48 \mu _{2,0} \mu _{2,3} \mu _{0,1}-96 \mu _{0,3} \mu _{1,0} \mu _{3,0} \mu _{0,1}-288 \mu _{0,2} \mu _{1,1} \mu _{3,0} \mu _{0,1}\nonumber\\
    &+32 \mu _{1,3} \mu _{3,0} \mu _{0,1}-288 \mu _{0,2} \mu _{1,0} \mu _{3,1} \mu _{0,1}+96 \mu _{1,2} \mu _{3,1} \mu _{0,1}
    +96 \mu _{1,1} \mu _{3,2} \mu _{0,1}
    +32 \mu _{1,0} \mu _{3,3} \mu _{0,1}\nonumber\\
    &+8 \mu _{0,3} \mu _{4,0} \mu _{0,1}+24 \mu _{0,2} \mu _{4,1} \mu _{0,1}-4 \mu _{4,3} \mu _{0,1}-360 \mu _{0,2}{}^2 \mu _{1,0}{}^4+24 \mu _{0,4} \mu _{1,0}{}^4-144 \mu _{1,1}{}^4\nonumber\\
    &+1728 \mu _{0,2} \mu _{1,0}{}^2 \mu _{1,1}{}^2-216 \mu _{1,0}{}^2 \mu _{1,2}{}^2-54 \mu _{0,2}{}^2 \mu _{2,0}{}^2+6 \mu _{0,4} \mu _{2,0}{}^2+72 \mu _{0,2} \mu _{2,1}{}^2-18 \mu _{2,2}{}^2\nonumber\\
    &+384 \mu _{0,3} \mu _{1,0}{}^3 \mu _{1,1}+576 \mu _{0,2} \mu _{1,0}{}^3 \mu _{1,2}-864 \mu _{1,0} \mu _{1,1}{}^2 \mu _{1,2}-288 \mu _{1,0}{}^2 \mu _{1,1} \mu _{1,3}-24 \mu _{1,0}{}^3 \mu _{1,4}\nonumber\\
    &+432 \mu _{0,2}{}^2 \mu _{1,0}{}^2 \mu _{2,0} -36 \mu _{0,4} \mu _{1,0}{}^2 \mu _{2,0}-432 \mu _{0,2} \mu _{1,1}{}^2 \mu _{2,0}+72 \mu _{1,2}{}^2 \mu _{2,0}
    -288 \mu _{0,3} \mu _{1,0} \mu _{1,1} \mu _{2,0}\nonumber\\
    &-432 \mu _{0,2} \mu _{1,0} \mu _{1,2} \mu _{2,0}+96 \mu _{1,1} \mu _{1,3} \mu _{2,0}+24 \mu _{1,0} \mu _{1,4} \mu _{2,0}-144 \mu _{0,3} \mu _{1,0}{}^2 \mu _{2,1}-864 \mu _{0,2} \mu _{1,0} \mu _{1,1} \mu _{2,1}\nonumber\\
    &+288 \mu _{1,1} \mu _{1,2} \mu _{2,1}+96 \mu _{1,0} \mu _{1,3} \mu _{2,1}+48 \mu _{0,3} \mu _{2,0} \mu _{2,1}-216 \mu _{0,2} \mu _{1,0}{}^2 \mu _{2,2}+144 \mu _{1,1}{}^2 \mu _{2,2}\nonumber\\
    &+144 \mu _{1,0} \mu _{1,2} \mu _{2,2}+72 \mu _{0,2} \mu _{2,0} \mu _{2,2}
    +96 \mu _{1,0} \mu _{1,1} \mu _{2,3}-24 \mu _{2,1} \mu _{2,3}+12 \mu _{1,0}{}^2 \mu _{2,4}-6 \mu _{2,0} \mu _{2,4}\nonumber\\
    &-72 \mu _{0,2}{}^2 \mu _{1,0} \mu _{3,0}+8 \mu _{0,4} \mu _{1,0} \mu _{3,0}+32 \mu _{0,3} \mu _{1,1} \mu _{3,0}+48 \mu _{0,2} \mu _{1,2} \mu _{3,0}-4 \mu _{1,4} \mu _{3,0}\nonumber\\
    &+32 \mu _{0,3} \mu _{1,0} \mu _{3,1}+96 \mu _{0,2} \mu _{1,1} \mu _{3,1}-16 \mu _{1,3} \mu _{3,1}+48 \mu _{0,2} \mu _{1,0} \mu _{3,2}-24 \mu _{1,2} \mu _{3,2}-16 \mu _{1,1} \mu _{3,3}\nonumber\\
    &-4 \mu _{1,0} \mu _{3,4}
    +6 \mu _{0,2}{}^2 \mu _{4,0}-\mu _{0,4} \mu _{4,0}-4 \mu _{0,3} \mu _{4,1}-6 \mu _{0,2} \mu _{4,2}+\mu _{4,4}
\end{align}

\bibliography{cumulant_aux,library}{}
\bibliographystyle{aasjournalv7}

\end{document}